\documentclass[useAMS,usenatbib]{mn2e}
\pdfoutput=1
\usepackage{times}
\usepackage{graphics,epsfig}
\usepackage{graphicx}
\usepackage{amsmath}
\usepackage{amssymb}

\newcommand{\kms}{km~s$^{-1}$}

\newcommand{\ha}{\ensuremath{{\rm H}\alpha}}

\newcommand{\ms}{\ensuremath{\rm M_{\odot}}}

\title[HI, star formation and feedback in low mass BCDs]{Atomic hydrogen, star formation and feedback in the lowest mass Blue Compact Dwarf galaxies}
\author[Roychowdhury et al.]{
Sambit Roychowdhury,$^{1}$\thanks{E-mail: sambit@ncra.tifr.res.in (SR); chengalu@ncra.tifr.res.in (JNC); kchibouc@gemini.edu (KC); ikar@sao.ru (IDK); tully@ifa.hawaii.edu (RBT); skai@sao.ru (SSK)} Jayaram N. Chengalur,$^{1\star}$ Kristin Chiboucas,$^{2,4\star}$
\newauthor Igor D. Karachentsev,$^{3\star}$ R. Brent Tully$^{4\star}$ and Serafim S. Kaisin$^{3\star}$\\
       \\ 
       $^{1}$NCRA-TIFR, Post Bag 3, Ganeshkhind, Pune 411 007, India\\
       $^{2}$Gemini Observatory, 670 North A'ohoku Place, Hilo, HI 96720, USA\\
       $^{3}$Special Astrophysical Observatory, Russian Academy of Sciences, N. Arkhyz, KChR 369167, Russia\\
       $^{4}$Institute for Astronomy, University of Hawaii, 2680 Woodlawn Drive, Honolulu, HI 96822, USA}

\begin{document}
\date{}

\pagerange{\pageref{firstpage}--\pageref{lastpage}} \pubyear{}

\maketitle

\label{firstpage}

\begin{abstract}
We present the results from a search for HI emission from a sample of newly discovered dwarf galaxies in the M81 group. HI is detected in three galaxies, all of which are classified as BCDs. The HI masses of these galaxies are $\sim 10^{6}$~\ms, making these some of the lowest mass BCDs known.  For these three galaxies FUV images (from GALEX) and \ha\ images (from the Russian 6m BTA telescope) are available.The \ha\ emission is very faint, and, in principle could be produced by a single O star. Further, in all cases we find offsets between the peak of the FUV emission and that of the \ha\ emission.  Offsets between the most recent sites of star formation (i.e. those traced by \ha) and the older  sites (i.e. those traced by FUV) would be natural if the star formation is stochastic. In spite of the expectation that the effects of mechanical feedback from star formation would be most directly seen in the smallest galaxies with low gravitational potentials, we only see tentative evidence of outflowing HI gas associated with the star forming region in one of the galaxies.
\end{abstract}


\section{Introduction}
\label{sec:int}

Accurate determination of  the faint end slope of the Luminosity Function in groups of galaxies is very important to resolve the discrepancy between the number of dwarf galaxies observed and that predicted by dark matter simulations of hierarchical structure formation \citep{kly99,moo99}, especially in low-density environments \citep{bar07,rob07}. An excellent candidate low-density environment group for carrying out a deep census of dwarfs is the M81 group of galaxies, for which it is possible to detect galaxies as faint as M$_R \sim -9$, and for which accurate TRGB distances can be determined for individual galaxies. \citet{chi09} present a census of the dwarfs in this group, determined from a large area survey using the Megacam instrument on the CFHT. A total of 22 candidate dwarf galaxies were found in this survey. Here we present follow up HI observations of some of these newly discovered dwarf galaxies.

The observations were done at the Giant Metrewave Radio Telescope (GMRT) and targeted a sub-sample of galaxies that showed evidence for active star formation, and as such was expected to have a significant HI content. At the redshift of the M81 group, HI emission from group members is expected to overlap in velocity with emission from HI in our Galaxy. An interferometric survey is hence the only reliable method for robustly detecting gas associated with the dwarfs in this group.
Deeper follow-up observations were obtained using the GMRT and the Westerbork Synthesis Radio telescope (WSRT) for the galaxies in which HI was detected.

The target galaxies were selected based on the original Megacam CFHT survey data. However, subsequently FUV data from the GALEX satellite as well as \ha\ data from the Russian 6m BTA telescope also became available. We present and analyse this data in conjunction with the HI data from the GMRT and the WSRT. The galaxies for which we do have GMRT detections have HI masses $\sim 10^{6}$ \ms, and star formation rates $\sim 10^{-4}$~\ms/yr. They represent some of the smallest mass star forming galaxies known. Feedback into the ISM from star formation is expected to be most pronounced for small galaxies, and we use our data to explore this issue too.

Below, in Sec.~\ref{sec:obsr} we present the sample, briefly describe the observations and results. The results are discussed in detail in Sec.~\ref{sec:res}, and the summary and conclusions presented in Sec.~\ref{sec:summary}.

\section{Sample, Observations and Results}
\label{sec:obsr}

\begin{table*}
\begin{center}
\caption{The sample}
\label{tab:samp}
\begin{tabular}{|lccccccccc|}
\hline
Name&$\alpha$ (J2000)&$\delta$ (J2000)&M${\rm{_{r'}}}^a$&a$\times$b$^b$&D${_{\rm Ho}}^c$&Type$^d$&D$^e$&v${\rm{_{rad}^{Sub}}}^f$&v${\rm{_{rad}^{BTA}}}^g$\\
~&(h~m~s)&($^\circ$~$^\prime$~$^{\prime\prime}$)&~&($^\prime~\times~^\prime$)&($^\prime$)&~&(Mpc)&(km s$^{-1}$)&(km s$^{-1}$)\\
\hline
\hline
d0926+70&09 26 27.9&+70 30 24&-9.7&0.8$\times$0.6&0.40&dI&3.3$^{+0.2}_{-0.2}$&&\\
d0944+71&09 44 34.4&+71 28 57&-12.5&1.3$\times$0.8&0.90&dI/dSph&3.4$^{+0.1}_{-0.1}$&&\\
d0958+66&09 58 48.5&+66 50 59&-13.2&1.6$\times$0.8&0.88&BCD&3.8$^{+0.1}_{-0.1}$&+33$\pm$94&+90$\pm$50\\
d1009+70&10 09 34.9&+70 32 55&~$^*$&0.6$\times$0.5&0.73&dI&~$^*$&&\\
d1012+64&10 12 48.4&+64 06 27&-13.4&1.4$\times$0.9&0.90&BCD&3.7$^{+0.1}_{-0.1}$&&+150$\pm$50\\
d1014+68&10 14 55.8&+68 45 27&-9.4&0.4$\times$0.4&0.20&dSph&3.8$^{+0.3}_{-0.3}$&&\\
d1028+70&10 28 39.7&+70 14 01&-12.4&1.4$\times$0.9&0.76&BCD&3.8$^{+0.1}_{-0.1}$&-90$\pm$79&-114$\pm$50\\
d1041+70&10 41 16.8&+70 09 03&-9.3&0.8$\times$0.4&0.48&dI&3.7$^{+0.2}_{-0.3}$&&\\
\hline
\hline
\end{tabular}
\end{center}
\begin{flushleft}
$^a$ estimated extinction corrected r${'}$ absolute magnitude, from \citet{chi09} corrected for the measured distance\\
$^b$ 28 mag/arcsec$^2$ diamaters of the Megacam images\\
$^c$ Holmberg diameter\\
$^d$ as listed in \citet{chi09}\\
$^e$ extinction corrected TRGB distances using HST, from \citet{chi12}\\
$^{f,g}$ Optically measured radial velocities from the Subaru telescope or 6m BTA telescope\\
$^*$ Not in M81 group, no distance measured.
\end{flushleft}
\end{table*}

\begin{table*}
\begin{center}
\caption{Observation Details}
\label{tab:obs}
\begin{tabular}{|lcccccccc|}
\hline
Name&Array&Date of Observation&Flux calibrator&Phase calibrator&Time on source&Channel width&Synthesised beam&Noise\\
&&&&&(hours)&(\kms)&(arcsec$^2$)&(mJy)\\
\hline
\hline
d0926+70&GMRT&04 Sep 07, 07 Sep 07&3C147,3C286&0834+555,1035+564&$\sim$6.9&3.3&33$\times$32&2.0\\
d0944+71&GMRT&29 Oct 07&3C147,3C286&0834+555&$\sim$6.1&3.3&38$\times$28&1.9\\
d0958+66&GMRT&04 Sep 07, 07 Sep 07&3C147,3C286&0834+555,1035+564&$\sim$6.0&3.3&34$\times$33&1.4\\
        &GMRT&01 Jan 12&3C48,3C286&0834+555,1035+564&$\sim$9.2&3.3&40$\times$40 , 12$\times$11&1.6 , 0.9\\
d1009+70&GMRT&04 Sep 07, 07 Sep 07&3C147,3C286&0834+555,1035+564&$\sim$6.1&3.3&35$\times$32&2.2\\
d1012+64&GMRT&28 Oct 07, 17 Jan 08&3C147,3C286&0834+555&$\sim$6.5&3.3&33$\times$30 , 13$\times$13&2.0 , 1.1\\
        &WSRT&16 Sep 11&3C48,3C286&3C48,3C286&$\sim$11.9&3.1&42$\times$42 , 23$\times$23&1.2 , 0.9\\
d1014+68&GMRT&26 Nov 07&3C147,3C286&0834+555&$\sim$7.1&3.3&48$\times$43&2.4\\
d1028+70&GMRT&28 Aug 07&3C286&1035+564&$\sim$3.1&3.3&39$\times$29&1.3\\
        &GMRT&02 Jan 12&3C48,3C286&1035+564,1313+675&$\sim$8.0&3.3&43$\times$43 , 14$\times$14&1.7 , 1.0\\
d1041+70&GMRT&28 Aug 07, 08 Sep 2007&3C147,3C286&0834+555&$\sim$9.3&3.3&22$\times$18&0.9\\
\hline
\hline
\end{tabular}
\end{center}
\end{table*}

\begin{table}
\begin{center}
\caption{Results from HI observations}
\label{tab:HIres}
\begin{tabular}{|lccc|}
\hline
Name& Velocity$^a$&Velocity$^b$&M$_{\rm HI}^c$\\
&(\kms)&(\kms)&(10$^6$\ms)\\
\hline
\hline
d0926+70&-208 to 567&&$<$ 0.31\\
d0944+71&-137 to 637&&$<$ 0.31\\
d0958+66&&89.6$\pm$0.9 , 29$\pm$2&0.73$\pm$0.08$^*$\\
d1009+70&-208 to 567&&-\\
d1012+64&&189.7$\pm$0.4 , 19$\pm$1&1.04$\pm$0.09$^*$\\
d1014+68&-139 to 636&&$<$ 0.48\\
d1028+70&&-68.9$\pm$0.5 , 19$\pm$1&1.3$\pm$0.1$^*$\\
d1041+70&-208 to 567&&$<$ 0.17\\
\hline
\hline
\end{tabular}
\end{center}
\begin{flushleft}
$^a$ velocity range searched in case of non detections\\
$^b$ V$_{\rm peak}$, $\Delta$V$_{\rm 50}$ in the case of detections\\
$^c$ 3$\sigma$ upper limits in the case of non detections\\
$^*$ error on mass includes flux and distance measurement errors.\\
\end{flushleft}
\end{table}

\begin{table*}
\begin{center}
\caption{Derived parameters for detected galaxies}
\label{tab:der}
\begin{tabular}{|lcccccccccc|}
\hline
Galaxy &D$_{\rm HI}$&i$_{\rm HI}$&M$_{\rm dyn}$&${\rm \frac{D_{HI}}{D_{Ho}}}$&M$_{\rm B_t}$&${\rm \frac{M_{HI}}{L_{B_t}}}$&SFR$_{\rm FUV}$&$\tau_{\rm HI}$&L$_{\rm H \alpha}$&SFR$_{\rm H \alpha}$\\
~&($^\prime$)&($^o$)&(10$^8~{\rm M_{\odot}}$)&&&(${\rm \frac{M_\odot}{L_{\odot B}}}$)&(${\rm M_{\odot}~yr^{-1}}$)&(Gyr)&(${\rm ergs~s^{-1}}$)&(${\rm M_{\odot}~yr^{-1}}$)\\
\hline
\hline
d0958+66&1.5&$\sim$29&$\sim$1.69&1.7&$-$11.90&0.08&5.0$\times {\rm 10^{-4}}$&1.5&3.7$\times {\rm 10^{36}}$&3.0$\times {\rm 10^{-5}}$\\
d1012+64&2.0&$\sim$25&$\sim$1.26&2.2&$-$12.24&0.09&6.2$\times {\rm 10^{-4}}$&1.8&2.45$\times {\rm 10^{37}}$&1.94$\times {\rm 10^{-4}}$\\
d1028+70&2.3&$\sim$44&$\sim$0.54&3.0&$-$11.70&0.17&7.0$\times {\rm 10^{-4}}$&1.8&3.03$\times {\rm 10^{37}}$&2.40$\times {\rm 10^{-4}}$\\
\hline
\hline
\end{tabular}
\end{center}
\end{table*}

\begin{figure*}
\begin{center}
\includegraphics[width=7truein]{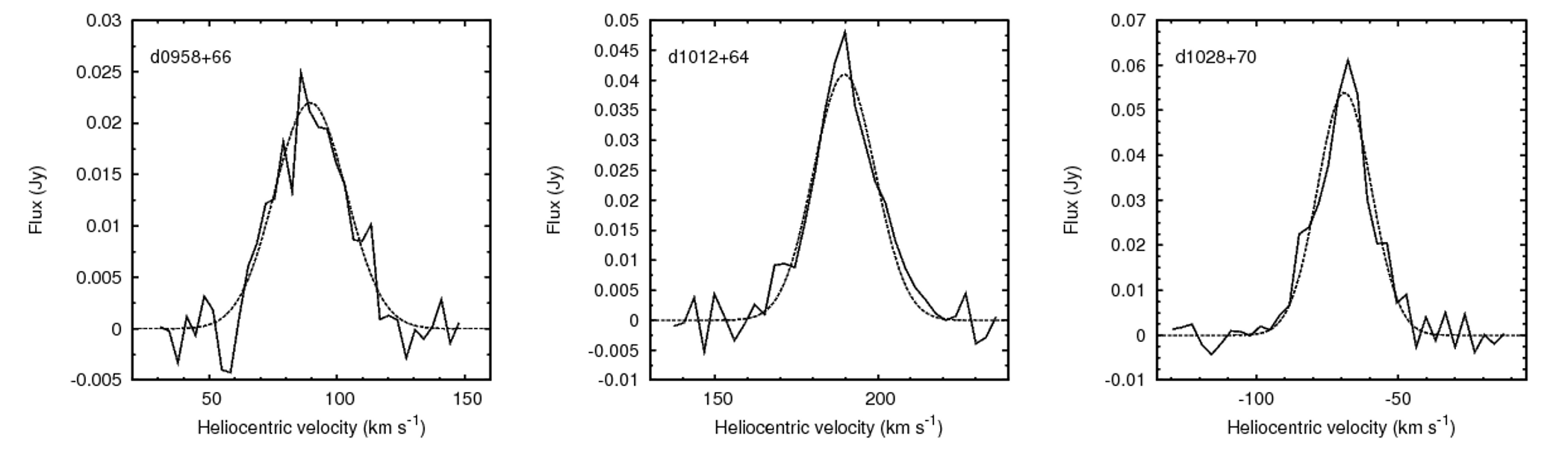}
\end{center}
\caption{HI spectra of the three detected BCDs derived using the follow-up observations, from the low resolution ($\sim 40^{''}$) data cubes. The single Gaussian fit to each spectrum is also shown as dashed lines.}
\label{fig:spec}
\end{figure*}

\begin{figure*}
\begin{center}
\includegraphics[width=7truein]{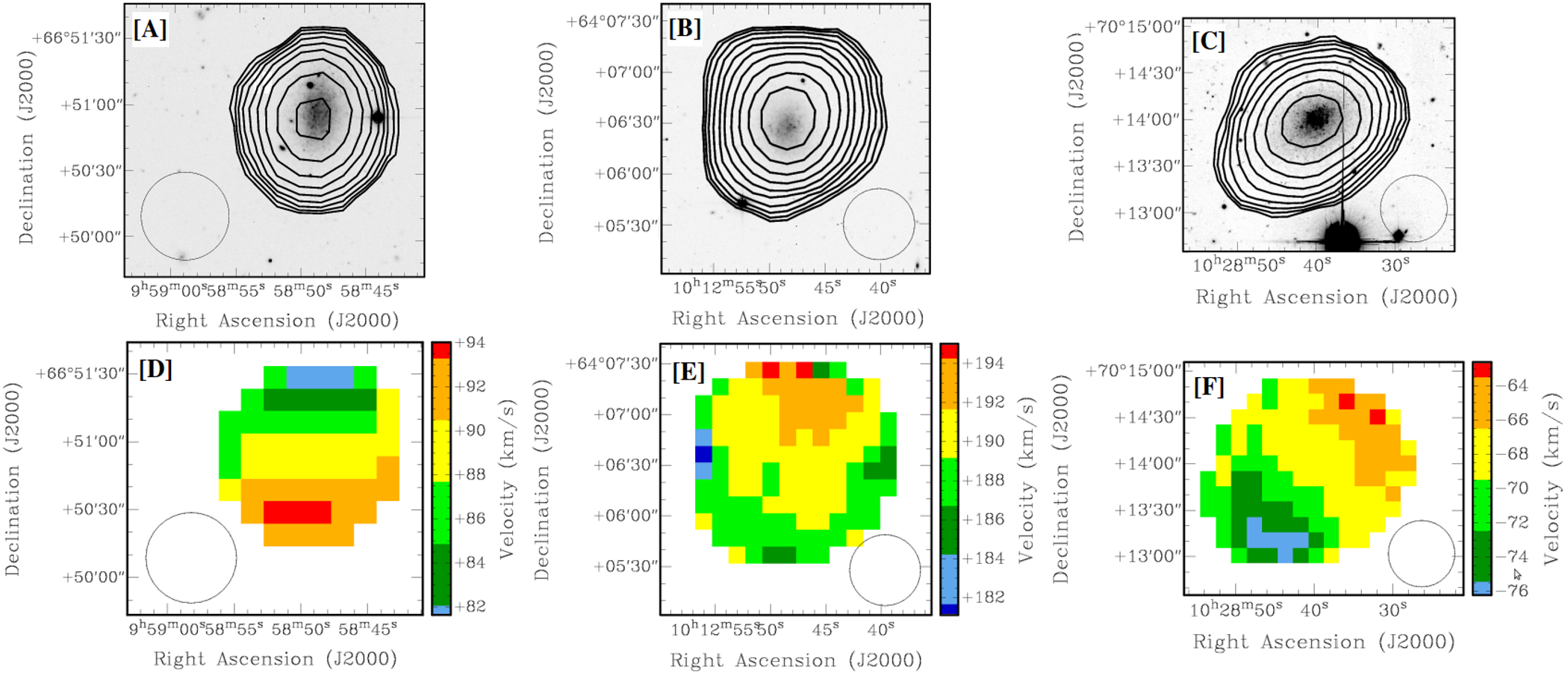}
\end{center}
\caption{The upper row shows the overlays of the low spatial resolution HI images (contours) of the BCDs d0958+66 [A], d1012+70 [B], and d1028+64 [C] on their respective R band Megacam images (greyscales). The resolutions are $40^{''}$, $42^{''}$ and $48^{''}$ for d0958+66, d1012+70 and d1028+64 respectively (beams also shown as circles in the images). The contours start at the 5$\sigma$ sensitivity levels for each galaxy, with each consecutive contour $\sqrt{2}$~times the previous one (please see Section~\ref{sec:obsr} for details about how moment maps were constructed).
The lower row shows the integrated velocity fields for the three galaxies (d0958+66 [D], d1012+70 [E] and d1028+64 [F]) at the same spatial resolution as the images in the upper row. For each galaxy, the images in the upper and lower rows are of the same spatial extent.}
\label{fig:uv5ov}
\end{figure*}

The sub sample chosen to be observed with the GMRT is listed in Table~\ref{tab:samp}. The main selection criterion was that the galaxy should have been classified as a dIrr or BCD, so that it was likely to contain HI. At the time of preparing the observing sample, d1014+68 was suspected to be a transition galaxy and was hence also added to the sample. Further, subsequent to the GMRT observations, d1009+70 was found not to belong to the M81 group and hence no distance for it was measured.

The details of the initial GMRT observations, as well as the later GMRT and WSRT observations of three galaxies in the sample, are given in Table~\ref{tab:obs}.
Please note that the noise value quoted in Column (9) is the rms noise in a line free channel of the (continuum subtracted) data cube made at the corresponding spatial resolution mentioned in Column (8).
 The data were reduced using the Astronomical Image Processing System (AIPS) using standard spectral line data analysis procedures.
Total intensity and velocity field maps were created using the AIPS task MOMNT, with the parameters in the adverb FUNCTYPE set to `HG' and the adverb CELLSIZE set to 3 6 (Except for the highest resolution moment maps of d1012+64 for which the parameters were set to 3 3).

Table~\ref{tab:HIres} gives a summary of the basic deductions one can make from the HI observations of these galaxies.
Mass limits based on the measured noise are provided for the non-detections.
These are 3$\sigma$~mass limits assuming the galaxy to be unresolved in a $\sim 40^{''}$~beam, and having a typical velocity  width of 20~\kms~\citep{beg08}.
HI was detected in three of the sample galaxies, and deeper observations of these three galaxies were later done with the GMRT and the WSRT.
We use these more sensitive observations for all subsequent analysis.
The peak and velocity spread of the detected HI line is given for these three galaxies, and the values are based on the low resolution data cubes.
For each galaxy, only channel fluxes more than two times the rms of the fluxes measured in line free channels were considered while obtaining the total flux.
As can be seen from the table, HI emission has been detected only from those galaxies that are classified as BCDs. The HI masses of the detections are all $\sim 10^6$\ms, making these the lowest HI mass BCDs with interferometric observations.
The integrated spectra for the detected galaxies are shown in Figure~\ref{fig:spec}. 

Figure~\ref{fig:uv5ov} shows an overlay of the low resolution ($\sim 40^{''}$) GMRT and WSRT HI images of the three detected BCDs and the Megacam R band images.
The moment 0 and moment 1 maps shown for all the galaxies were constructed using the previously explained method after applying a flux threshold of 2 times the noise in the line free channels.
It is apparent that the HI discs extend much more than the stellar disc (see Column (5) in Table~\ref{tab:der}), and the stellar disc is approximately coincident with the central density peak of HI for all the galaxies.
Looking at the velocity fields for the galaxies, we see that in all cases there is an overall velocity gradient approximately aligned with the morphological major axis. However, the isovelocity contours are not regular, and the velocity fields cannot be modelled as simple rotating disks. Not surprisingly, attempts to fit tilted ring models to the velocity fields did not lead to meaningful fit parameters.

Table~\ref{tab:der}~lists various measured global parameters of the three detected BCDs. The entries in the table are:
Column(1)~galaxy name,
Column(2)~HI diameter D$_{\rm HI}$, from the ${\rm 10^{19}~cm^{-2}}$ column density contour measured in the low resolution Moment 0 map (shown in Figure~\ref{fig:uv5ov}). For all the three galaxies, this contour was near or above the 3$\sigma$ sensitivity contour.
Column(3)~an indicative inclination angle ($i$) of the HI disk. The inclination angle was determined by measuring the axial ratio of the above mentioned ${\rm 10^{19}~cm^{-2}}$ column density isophote and assuming an intrinsic axial ratio of 0.57 \citep{roy10} if one assumes that the HI discs are oblate spheroids. The inclination angle estimated in such a manner is highly uncertain, and is only used here to quantify what is apparent from looking at the HI discs of the galaxies: d0958+66 and d1012+64 have very face-on HI discs and d1028+70 has a slightly inclined disc.
Column(4) an indicative dynamical mass estimated as $ M_{\rm dyn} = (\Delta V_{50}/2\sin(i))^2 \times (D_{\rm HI}/2)/G $.
Column(5) the ratio of the HI to the Holmberg diameter.
Column(6) is the absolute blue magnitude obtained by applying the distance measurements of the galaxies to the blue magnitudes estimated by eye from POSS-II images tabulated in \citet{chi09}.
Column(7) the HI mass-to-light ratios of the three galaxies, using the magnitudes tabulated in column (6). 
Column(8) gives the total SFR as deduced from FUV emission (SRF$_{\rm FUV}$) as measured by \emph{GALEX}, and using the calibration given in \citet{ken98} (more details on this are given below).
Column(9) The gas consumption timescale $\tau_{\rm HI}$ defined as $\tau_{\rm HI} =  {\rm M}_{\rm HI}/{\rm SFR}^{\rm FUV}$.
Column(10) The total \ha~luminosity as determined from the 6m BTA telescope.
Column(11) the SFR computed from the \ha\ luminosity, using the calibration in \citet{ken98} (more details on the last column are also given below).

\citet{gil03} suggest a  set of quantitative criteria for classifying a galaxy as a BCD. Unfortunately these involve photometry in the B, R and K bands, which is not available for our galaxies. However from the measured \ha\ luminosity and M$_{\rm B}$ for our galaxies we see that they follow the same general trend as the brighter BCDs in \citet{gil03} (see their Fig. 4e).
Also, the gas consumption timescales for our galaxies is similar to that typical for brighter BCDs \citep{san08}.
It is however interesting to note that the relative content of neutral hydrogen in these galaxies is lower than that seen for brighter BCDs with HI observations, more so keeping their faint blue magnitudes in mind \citep[e.g., comparing with Figure 2 in ][the ratios are all below the best fit to their data]{huc05}.

The SFR from the FUV was computed by taking the total FUV flux contained within the best fit ellipse to the 28th magnitude/arcsec$^2$ isophote of the CFHT Megacam images. The centre and principal axes of this ellipse are given in Table~\ref{tab:samp}. This FUV flux was then corrected for the local extinction expected from dust in our galaxy.  Note that the presence of foreground galactic cirrus with substantial fine-scale structure in the line of sight towards the M81 group makes this correction somewhat uncertain \citep{sandage76,dav10}. No correction was made for extinction internal to the BCDs, since this is expected to be small.
The corrected FUV flux was converted to a SFR using the standard calibrations for obtaining SFRs given in \citet{ken98}. Note that this calibration assumes solar metallicity and a constant star formation rate over the last $\sim 100$ million years. From the luminosity-metallicity relation one would expect that our sample galaxies would have a lower than solar metallicity (see below), and further that since they are BCDs they may be undergoing a recent starburst. However, in view of the fact that there is some uncertainty in the foreground FUV extinction, no correction was made for these two effects. Similarly, the SFR was computed from the \ha\ flux assuming the standard calibration from \citet{ken98}. As discussed further below, this is unlikely to be a reliable estimate for the SFR in these galaxies.

\section{Discussion}
\label{sec:res}

\begin{table*}
\begin{center}
\caption{Revised SFRs from IGIMF theory}
\label{tab:ig}
\begin{tabular}{|lcccc|}
\hline
Galaxy &SFR$_{\rm FUV}^{\rm std}$&SFR$_{\rm FUV}^{\rm min1}$&SFR$_{\rm H \alpha}^{\rm std}$&SFR$_{\rm H \alpha}^{\rm min1}$\\
~&(${\rm M_{\odot}~yr^{-1}}$)&(${\rm M_{\odot}~yr^{-1}}$)&(${\rm M_{\odot}~yr^{-1}}$)&(${\rm M_{\odot}~yr^{-1}}$)\\
\hline
\hline
d0958+66&1.7$\times {\rm 10^{-3}}$&9.1$\times {\rm 10^{-4}}$&1.7$\times {\rm 10^{-3}}$&9.9$\times {\rm 10^{-4}}$\\
d1012+64&2.0$\times {\rm 10^{-3}}$&1.1$\times {\rm 10^{-3}}$&4.3$\times {\rm 10^{-3}}$&2.3$\times {\rm 10^{-3}}$\\
d1028+70&2.2$\times {\rm 10^{-3}}$&1.2$\times {\rm 10^{-3}}$&4.8$\times {\rm 10^{-3}}$&2.5$\times {\rm 10^{-3}}$\\
\hline
\hline
\end{tabular}
\end{center}
\end{table*}

\begin{figure*}
\begin{center}
\includegraphics[width=7truein]{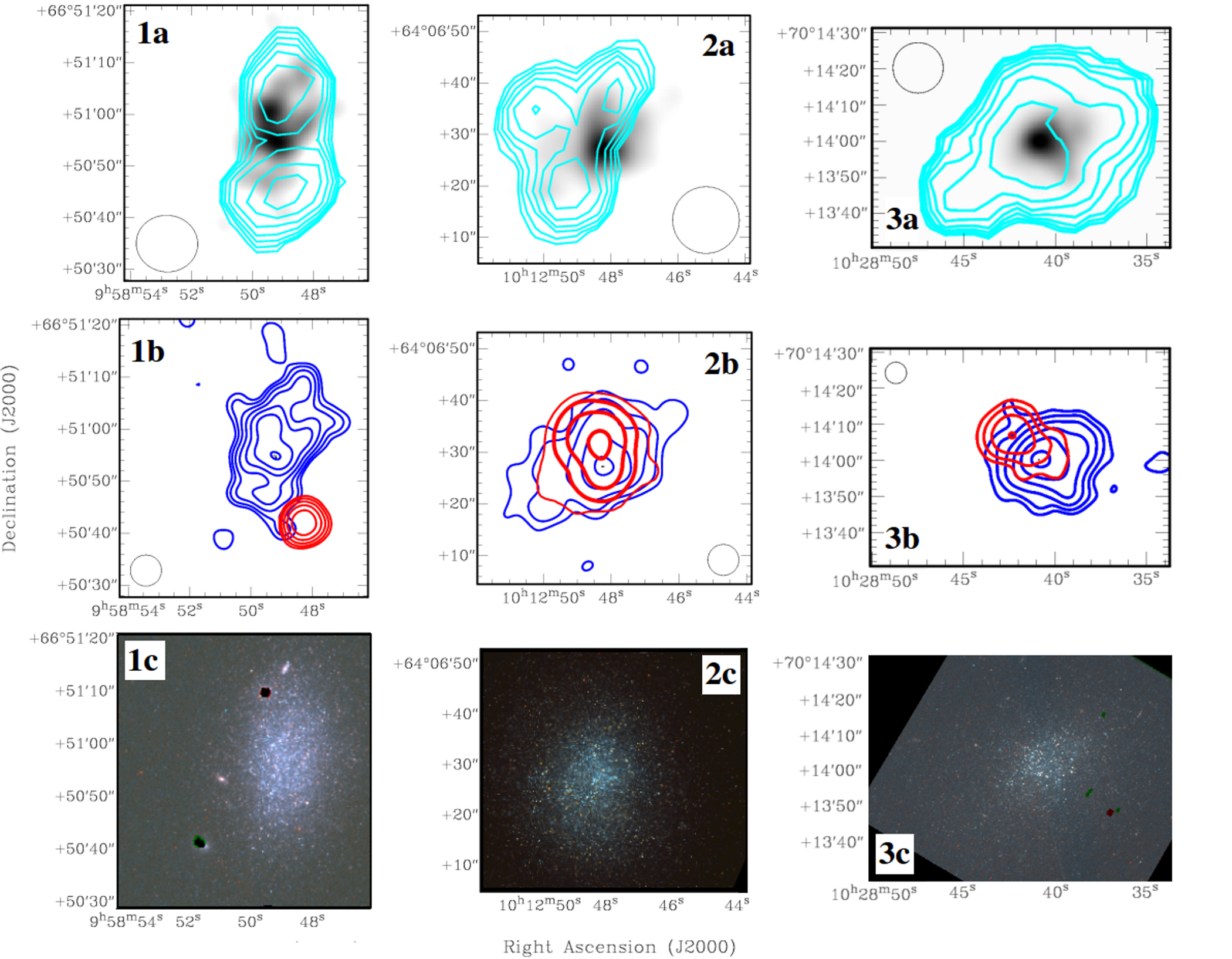}
\end{center}
\caption{Comparison of HI, FUV, \ha, and optical data for the BCDs d0958+66 [1], d1012+70 [2], and d1028+64 [3]. The first row [a] shows the highest spatial resolution HI contours in cyan (with beam sizes $12^{''}\times11^{''}$, $13^{''}\times13^{''}$ and $14^{''}\times14^{''}$ for d0958+66, d1012+70 and d1028+64 respectively, which are also shown as circles) overlayed on the FUV emission smoothed to $6^{''}$ resolution shown in greyscale. The contours start at the 3$\sigma$ sensitivity levels for each galaxy (please see Section~\ref{sec:obsr} for details about how moment maps were constructed) and subsequent contours are at intervals of $\sqrt{2}$. The second row [b] shows \ha~emission contours in red overlayed with FUV emission contours in blues (both smoothed to $6^{''}$ resolution, beam sizes also shown as circles). For d0958+66, the contours start at the 3$\sigma$ sensitivity levels for FUV and \ha~emission, and are spaced apart by $\sqrt{2}$. For d1012+70 and d1028+64, the contours start at the 5$\sigma$ sensitivity levels for FUV and \ha~emission, and are spaced apart by $2$. The greyscale images of FUV in row [a] start at the level of the first blue contours in row [b]. The lower row [c] shows the RGB images constructed from HST observations, for the same spatial region as the images in the top row, separately for each galaxy. In the RGB images, the red and green channels are the I (F814W) and V (F606W) band images, whereas the blue channel image is constructed as 2$\times$V-I. For d0958+66 and d1028+70 the images are from WFPC2, whereas for d1012+64 the images are from ACS.}
\label{fig:sfr_gas}
\end{figure*}

\begin{figure*}
\begin{center}
\includegraphics[width=6truein]{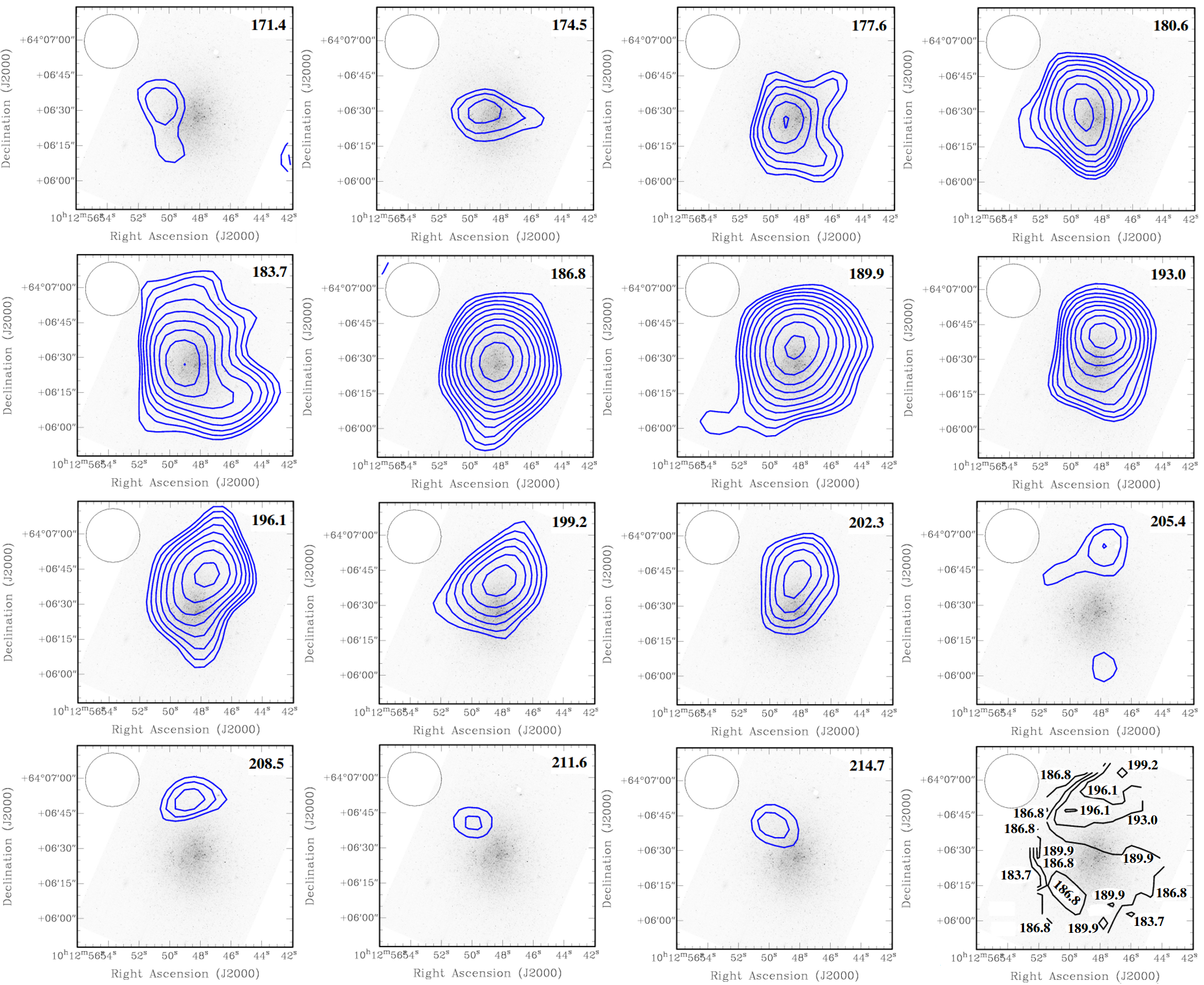}
\end{center}
\caption{The HI emission in different velocity channels of d1012+64, imaged at a spatial resolution of $23^{''}$ (shown as circles), is shown as blue contours. All cutouts represent the same spatial area with the central heliocentric velocity for each channel mentioned in the top right corner (in \kms). The HI (blue) contours start at the 3$\sigma$ sensitivity level, and are spaced apart by $1.2$. The HST ACS V (F606W) band image is shown as greyscale in all images, and represents the stellar disc. The grey contours shown in the last image is the integrated velocity field of the galaxy at the same resolution at which the HI emission for each channel is shown (please see Section~\ref{sec:obsr} for details about how moment maps were constructed). The velocity which corresponds to any particular contour is mentioned (in \kms).}
\label{fig:chmap}
\end{figure*} 

\begin{figure*}
\begin{center}
\includegraphics[width=6truein]{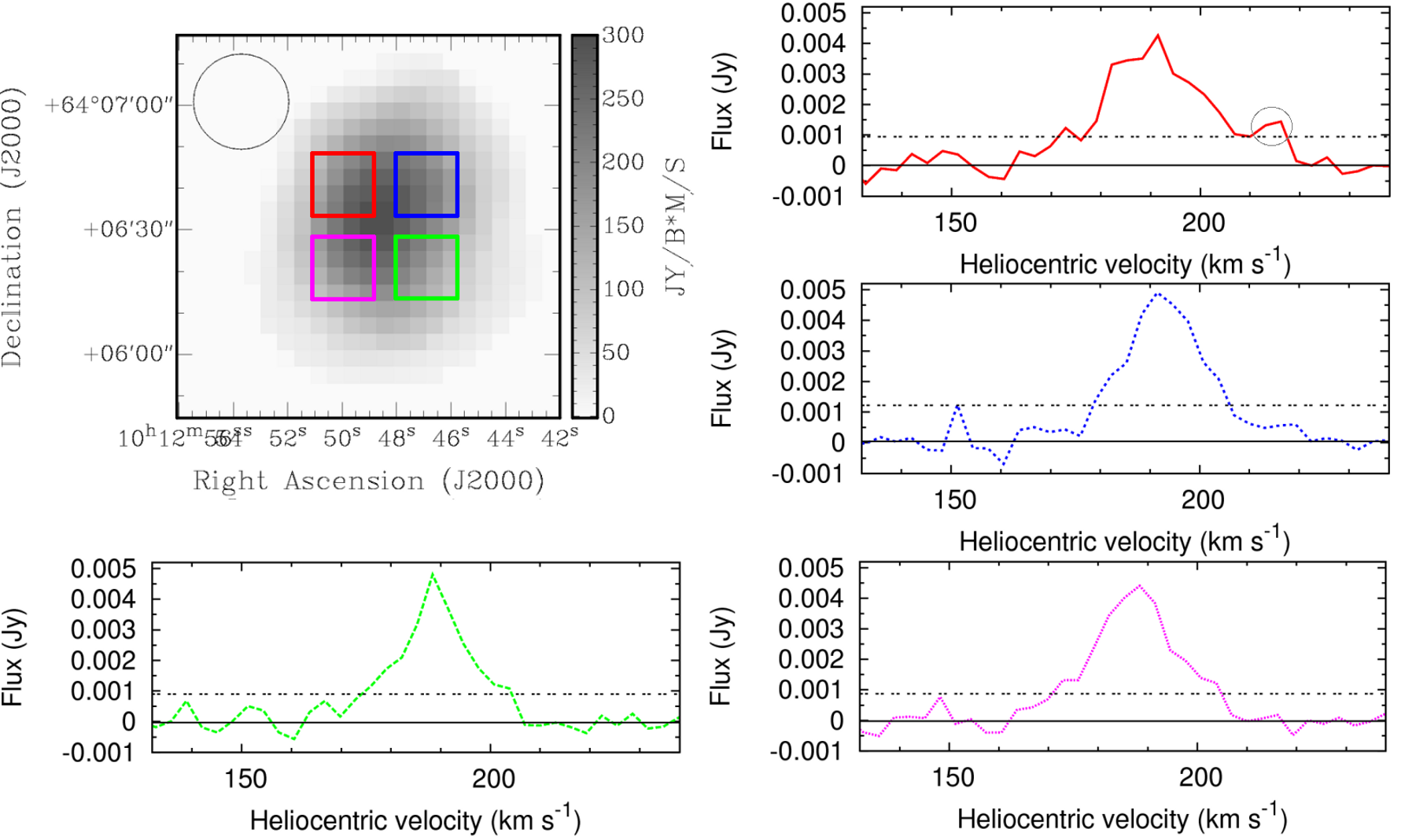}
\end{center}
\caption{Separate spectra for four different regions in the galaxy d1012+64. In the top left hand corner the moment 0 map in HI is shown in greyscale, with the four regions marked in different colours. The spatial extent of the image and the resolution at which HI has been mapped (also shown as a circle) are identical to the images shown in Figure~\ref{fig:chmap}. The spectra for each of the four regions are shown in the same colour in which the regions have been marked. The dashed line in each spectrum is at 3 times the rms noise of the fluxes in the line free channels for that particular spectrum. The red spectrum is for the region which shows the suspected HI outflow, marked here by the grey circle around the concerned channels.}
\label{fig:spreg}
\end{figure*}

\subsection{HI non-detections}

The newly-discovered dwarf galaxies in the M81 group \citep{chi09} chosen for follow-up HI observations were the brightest ones, and were thought to be star forming galaxies as complexes of young blue stars seemed to be present in all of them on initial visual inspection.
Subsequent Hubble Space Telescope (HST) observations revealed that d1009+70 is a background galaxy not in group (possible companion to UGC 5423 at a distance of 8.7 Mpc) and hence its HI emission may be too faint to be detected.
Of the other five galaxies with no detected HI emission, follow-up photometry using HST data has revealed that d0944+71 (initially could not unambiguously be classified as either dI or dSph), d1014+68 (initially thought to be transition type dwarf) and  d1041+70 (initially classified as dI as mentioned in Table~\ref{tab:samp}) are all dSph galaxies.
Even if the above galaxies have young stellar complexes, these are presumably small (as can be said from their faintness) and are being fed by HI reservoirs with masses below the detection limit of the present observation.
Follow-up HST photometry also shows that d0926+70 is indeed a dI galaxy, though with very little star formation which explains the non-detection of HI using similar reasoning as stated above for the dSph galaxies.  
For all the non-detections, there is the possibility that they have systemic velocities outside the velocity range searched.
But this is an unlikely possibility considering the fact that TRGB distance measurements constrain them to be in the M81 group, and the measured velocities of other M81 group members \citep{kar02} are in the range of velocities searched.

\subsection {H$\alpha$ and FUV emission}

In Figure~\ref{fig:sfr_gas}, the \ha~and FUV emission are shown with respect to the corresponding high resolution HI maps for the three BCDs.
The moment 0 maps shown for all the galaxies were constructed using the previously explained method after applying a flux threshold of 2 times the noise in the line free channels for d0958+66 and d1028+70, and 1.5 times the noise in the line free channels for d1012+64 (for which old GMRT data was used to make the high resolution map as WSRT cannot map at this resolution).
Also shown are the RGB images of the three galaxies constructed from HST data\footnote{The HST images and other materials for the three BCDs can be found at http://edd.ifa.hawaii.edu. Select the catalog CMDs/TRGB and then select the desired object in the column Name/CMD. Alternate names for the galaxies are d0958+66 = PGC0028826, d1012+64 = PGC0029735, d1028+70 = PGC5056941.}. 
A few things are apparent from the figures:
(i) although the star forming disc is at the central peak of the HI profile when the HI is imaged at a low resolution (Figure~\ref{fig:uv5ov}), the peaks of FUV and \ha~emission don't concur with the peaks of the HI emission when HI is images at high resolution.
(ii) There are offsets between the peaks of the \ha\ emission and that 
of the FUV emission, these are most pronounced in d0958+66 and d1028+70.
(iii) The stellar disc as appears in the RGB images is better traced by the FUV emission.

Before examining the differences between the \ha\ and the FUV emission it is useful to take a look at the total integrated fluxes from the entire galaxy (Table~\ref{tab:der}). As can be seen the total \ha\ fluxes are very low -- in fact they are consistent with that expected from a {\it single} O star. The ionized flux of O stars is critically dependent on the metallicity. Since we do not have abundance quality optical spectra to determine the metallicity, the metallicity of our galaxies can be estimated from the L-Z relation.  The absolute blue magnitudes of our BCDs are $\sim  -12.0$ mag. From the L-Z relation for BCDs \citep[see][]{ekt10},  the expected metallicity is 12+log(O/H) = 7.30, i.e. about 0.05 times the solar metallicity. From the 0.05 solar metallicity model atmospheres given in \citet{smith02}, the \ha\ flux expected from an ionization bounded HII region excited by a single O4 star is $3.4\times 10^{37}$ ergs/s, the corresponding number for an O8 star is $3.4\times 10^{36}$ ergs/s. These values bracket the measured \ha\ fluxes for our galaxies. It is worthwhile to note that the L-Z relation for BCDs used here is quite well constrained, as it leaves out the deviant extremely metal deficient (XMD) BCDs (see their Figure 2). Even if the galaxies considered here turn out to be XMD BCDs, then the fluxes expected from the HII regions will only increase.

From the SFRs deduced using the total FUV and \ha\ fluxes listed in Table~\ref{tab:der} one can also see that for all the three BCDs the deduced SFR$^{\rm FUV}$ is $>$ SFR$^{\rm H\alpha}$. This is typical for low mass galaxies \citep{lee09,hun10,roy11}. As discussed in detail in these (and other papers), this is likely due to the fact that at low star formation rates, the number of massive stars that are present at any given time is subject to substantial statistical fluctuations. The \ha\ emission (which comes from massive M$>$ 17~M$_{\odot}$ stars) will in general be weaker than that predicted from the FUV luminosity (which comes from intermediate M $>$ 3~M$_\odot$ stars).

A lower \ha\ flux compared to what would be expected from the observed FUV 
emission and a standard Salpeter IMF is also predicted by the  integrated 
galaxial initial mass function (IGIMF) theory \citep{wei05}. In this theory, 
the clustered nature of star formation, along with the assumption that the most
massive star formed in a cluster depends on the total mass of the cluster,
results in the total sampled IMF to differ from the Salpeter IMF. The theory
predicts that the FUV and \ha~luminosities decrease non-linearly with the total
SFR; the effect is most pronounced for galaxies with low SFRs. \citet{pfl09a} 
show that recalibrations using the IGIMF explain to a large extent the 
drastic decrease in \ha~luminosities for dwarf galaxies. Table~\ref{tab:ig} 
shows the recalibrated SFRs for our BCDs, from both FUV and \ha, for two 
sets of IGIMF models (i.e. those called standard and minimal-1 in 
\citet{pfl07}) which cover the full range of IGIMF models. Columns(2) and (3) 
are the recalibrated SFRs from FUV using calibrations given in 
\citet{pfl09b}, while Columns(4) and (5) are the recalibrated SFRs 
from \ha~using calibrations given in \citet{pfl07}. The recalibrated SFR 
estimates match quite well for d0958+66, but for the other two BCDs, 
the SFR as derived from \ha~is approximately twice the SFR estimated using 
FUV flux. From the above discussion, it is not clear whether there is any way to distinguish between the effects of stochasticity in high mass star formation and variable IMF on the \ha~and FUV SFRs.
But recent modelling using spectral synthesis codes of star clusters which include stochastic sampling of the IMF \citep[e.g., ][]{fum11,eld12} suggest that in the low SFR regime it is the determining factor.
The data presented in \citet{fum11} (see their Figure 1) is for starbursts with slightly higher FUV fluxes than in the current case, however our data is consistent with the trends shown in their figure. As those authors emphasize, stochastic sampling of the IMF could occasionally result in small clusters with a disproportionate number of massive stars (which otherwise would be evidence for a ``top heavy'' IMF), and this appears to be the case here, especially for the galaxy d0958+66.

\subsection{Signs of feedback from star formation}

We expect that the effects of mechanical feedback from star formation into the ISM would be more pronounced and easier to detect in smaller low mass galaxies.
For example, \citet{mac99} modelled the effects of supernova explosions from starbursts on the interstellar medium of dwarf galaxies to show that for galaxies with total gas mass $\le~{\rm 10^7}$~\ms\ a significant fraction of the ISM could be expelled. This result however is sensitive to both the concentration of the starburst and the geometry of the ISM. 
Feedback from star formation is also believed to play an important role in determining the final dark matter distribution. 
For example, \citet{gov10} show that supernova driven outflow of low angular momentum gas from the centres of star forming galaxies inhibit the formation of bulges and can account for the shallow (instead of cuspy) central dark matter profiles seen for nearby gas rich dwarfs. This weakening of the central dark matter cusp occurs even if there is no escape of baryonic matter from the galaxy. On the observational front, \citet{vaneymeren10} as well as \citet{cannon11} both find evidence for feedback from star formation into the neutral ISM of dwarf galaxies. 
But even though our galaxies represent some of the smallest mass, actively star forming systems, evidence for feedback between the star formation on the neutral gas distribution in two of them (d0958+66 and d1028+70) is not clearly evident. 

The only evidence we see of mechanical feedback form star formation on the HI distribution is in the galaxy d1012+64.
Figure~\ref{fig:chmap} shows the HI channels maps for this galaxy at an intermediate resolution (chosen to optimise between localising the HI emission to a particular region and having enough signal-to-noise ratio to pick up the emission).
The stellar disc, and the integrated velocity field at the same resolution (moment 1 map) are shown for reference. 
The velocity field shows a South-East to North alignment as one moves forward in the velocity space.
The channels of interest are the last two ones shown in the Figure, with central heliocentric velocities of 211.5 \kms~and 214.7 \kms.
Here the HI suddenly moves out of the South-East to North flow for increasing velocity, and returns to the central region of the HI disc of the galaxy. 
The noteworthy fact is that the HI emission in these two channels is right at the edge of the stellar disc, and overlapping with the \ha~emission coming from recent star formation (compare with Figure 3, 2b and 2c).
To confirm that there is high velocity HI concentrated in only one region, we plot the HI spectrum for four different regions around the peak of HI emission in Figure~\ref{fig:spreg}.
There is statistically significant detection of HI in two channels at heliocentric velocities above 210 \kms~in only the region to the North-East, as is seen from Figure~\ref{fig:chmap} too.
The total mass of HI in the relevant channels is $\sim{\rm 1.4~\times~10^4}\ms$.
Taking the mean velocity difference between the channel velocities and the systematic velocity of HI ( $\sim{\rm 23.5}$~\kms), the estimated kinetic energy being carried away by the gas is hence $E_{KE} \sim 8~\times~10^{49}$ ergs.
Considering the wind mechanical energy output from an O8 star within a typical lifetime of $\sim 3$~Myr is $\sim 1.1~\times~10^{48}$~ergs \citep{smith02}, just the mechanical energy output from the \ha~emitting cluster in d1012+64 may not be enough to produce the gas outflow seen.
But this much of energy can easily be accounted for even by the typical energy output of a single supernova (i.e. $\sim 10^{51}$~erg).
Hence, what we see is most probably signature of neutral gas outflow associated with the star forming region of d1012+64.

\section{Summary and Conclusions}
\label{sec:summary}

Our search for HI in the newly discovered dwarf galaxies in the M81 group has resulted in three new detections. The detected galaxies are some of the lowest mass BCDs known, with star formation rates $\sim 10^{-4}$ \ms/yr. The star formation can be traced by both \ha\ and FUV emission, and as is the case for most other dwarfs, the observed \ha\ flux is less than what would be predicted from the star formation rate deduced from the FUV. This is likely to be a result of stochastic sampling of the upper end of the mass function. The \ha\ emission is also offset from the peak of the FUV emission. We find tentative evidence of HI gas outflow at the edge of the stellar disc of one of the galaxies. Even though this result agrees with the expectation that the effects of mechanical feedback from star formation would be most directly seen in the smallest galaxies, surprisingly no clear evidence of feedback associated with star formation is seen in the HI discs of the other two galaxies.

\section*{Acknowledgments}
We thank the GMRT staff for having made possible the observations used 
in this paper. The GMRT is run by the National Centre for Radio 
Astrophysics of the Tata Institute of Fundamental Research.
The Westerbork Synthesis Radio Telescope is operated by the ASTRON (Netherlands Institute for Radio Astronomy) with support from the Netherlands Foundation for Scientific Research (NWO).
RBT and KC acknowledge support from NSF award AST03-07706 and NASA/STScI award HST-GO-11126.
IDK and SSK were supported in part by the RFBR grants 10-02-92650 and 10-02-00123.
Some of the data presented in this paper were obtained from the 
Multimission Archive at the Space Telescope Science Institute (MAST). 
STScI is operated by the Association of Universities for Research in 
Astronomy, Inc., under NASA contract NAS5-26555. Support for MAST 
for non-HST data is provided by the NASA Office of Space Science via 
grant NAG5-7584 and by other grants and contracts.

\noindent
We want to thank the anonymous referee for the detailed and constructive comments which helped improve the manuscript.

\bsp

\label{lastpage}

\end{document}